%% file: paper_index.tex
\documentclass{article}

\usepackage{mathtools}
\usepackage{hyperref}
\usepackage{graphicx}
\usepackage{pdfpages}
\usepackage{makecell}
\usepackage{subcaption}
\usepackage{rotating}
\usepackage{multirow}
\newcommand{\ourtool}{LB4MPI}
\usepackage[vlined,linesnumberedhidden,ruled,resetcount]{algorithm2e}
\SetAlgorithmName{Listing}{algorithmautorefname}{list of algorithms name}

\DeclarePairedDelimiter{\ceil}{\lceil}{\rceil}
\DeclarePairedDelimiter{\floor}{\lfloor}{\rfloor}

\newcommand{\Eq}{Eq.}

\begin{document}
\title{A Distributed Chunk Calculation Approach for Self-scheduling of Parallel Applications on Distributed-memory Systems}

\author{Ahmed Eleliemy and Florina M. Ciorba\\Department of Mathematics and Computer Science\\University of Basel, Switzerland}
\date{}
\maketitle
\clearpage
\tableofcontents
\newpage

\input{0.tex}
\input{1.tex}
\input{2.tex}

\input{3.tex}
\input{4.tex}
\input{5.tex}
\input{6.tex}

\input{7.tex}

\input{8.tex}

\section*{Acknowledgment}
This work has been supported by the Swiss National Science Foundation in the context of the Multi-level Scheduling in Large Scale High Performance Computers” (MLS) grant number 169123 and by the Swiss Platform for Advanced Scientific Computing (PASC) project SPH-EXA: Optimizing Smooth Particle Hydrodynamics for Exascale Computing. 

\bibliographystyle{elsarticle-num}
\bibliography{paper_references}

\end{document}

%% file: 0.tex
\begin{abstract}
		Loop scheduling techniques aim to achieve load-balanced executions of scientific applications. 
	Dynamic loop self-scheduling (DLS) libraries for \mbox{distributed-memory} systems are typically \mbox{MPI-based} and
	employ a centralized chunk calculation approach~(CCA) to assign variably-sized chunks of loop iterations. 
	We present a distributed chunk calculation approach~(DCA) that supports various types of DLS techniques.
	Using both CCA and DCA, twelve DLS techniques are implemented and evaluated in different CPU slowdown scenarios.
	The results show that the DLS techniques implemented using DCA outperform their corresponding ones implemented with CCA, especially in extreme system slowdown scenarios.

\end{abstract}
	
\paragraph*{Keywords} Dynamic loop self-scheduling~(DLS), Load balancing, Centralized chunk calculation, distributed chunk calculation
\newpage
	

%% file: 1.tex
\section{Introduction}
\label{sec:intro}
Loops are the prime source of parallelism in scientific applications~\cite{fang1990dynamic}.
Such loops are often irregular and a balanced execution of the loop iterations is critical for achieving high performance. 
However, several factors may lead to an imbalanced load execution, such as problem characteristics, algorithmic, and systemic variations. 
Dynamic loop \mbox{self-scheduling} (DLS) techniques are devised to mitigate these factors, and consequently, improve application performance.
The DLS's dynamic aspect refers to assigning independent loop iterations during applications' execution. 
The self-scheduling aspect means that processing elements (PEs) drive the scheduling process by requesting work once they become \textit{free}.
Both these aspects, \textit{dynamic} and \textit{\mbox{self-scheduling}}, make DLS techniques an excellent candidate to minimize loops' execution time and  achieve a balanced execution of scientific applications on parallel systems.

DLS techniques typically distinguish (1)~How many loop iterations to assign to individual PEs? and (2)~Which loop iterations to assign?
DLS techniques assign chunks of loop iterations to each \textit{free} and \textit{available} PE.
The calculation of the chunk size,  referred to as \textbf{chunk calculation}, is determined for each technique by a \textit{mathematical formula}.
DLS techniques typically \textit{assume no dependencies between loop iterations}, and therefore, loop iterations can be assigned and executed in any order.
However, DLS techniques also assume \textit{a central work queue}. 
PEs synchronize their accesses to the central work queue to avoid any overlap in the \textbf{chunk assignment}.
If a specific  DLS technique calculates two chunks of fifty  and  ten loop iterations for two PEs: P1 and P2, respectively,
both PEs need to synchronize their accesses to the central queue to ensure that the fifty loop iterations of P1 do not overlap with the ten loop iterations of P2.
The chunk assignment requires exclusive access to the central work queue. 
This exclusiveness means that if P2 obtains the access before P1, P2 will obtain the first ten loop iterations and leave the next fifty loop iterations to P1. 

There are two approaches to synchronize the chunk assignment:
(1)~Making only one PE responsible for accessing the central work queue on behalf of all other PEs.
(2)~Serializing the PEs' accesses to the central work queue.
Earlier DLS techniques such as guided \mbox{self-scheduling}~(GSS)~\cite{GSS}, and factoring~FAC~\cite{FAC}, were devised for \mbox{shared-memory} systems.
Thus, both synchronization approaches above were possible to be implemented.
In the first approach, one thread acts like a master that is exclusively permitted to access the work queue, while other threads act as workers and only request chunks of work.
In contrast, the second approach would involve using critical regions and atomic operations to safely access the central work queue.

In the middle of the 1990s, \mbox{distributed-memory} systems, such as clustered computational workstations, started to be a dominant architecture for high performance computing~(HPC) systems~\cite{BEOWULF,castagnera1994clustered}.
For these systems, having a single PE responsible for the chunk assignment is the \textit{only} available implementation approach.
Hence, the \mbox{master-worker} execution model has been a  prominent approach to implement DLS techniques on \mbox{distributed-memory} systems.
In the \mbox{master-worker} execution model, the master is a central entity that performs both  the chunk calculation and the chunk assignment.
This centralization may render the master a potential performance bottleneck in different scenarios.
For instance, the master degrade the performance of the entire application,  when it experiences a certain slowdown in its processing capabilities.

Although centralizing the chunk assignment does not mean centralizing the chunk calculation, many of the recent DLS techniques employ a \mbox{master-worker} execution model that centralizes both the chunk calculation and the chunk assignment at the master side~\cite{TFSS,ISS,DHTSS,AWF,AWFBC}. 
The current work extends our earlier distributed chunk calculation approach (DCA)~\cite{DCA} and makes the following unique contributions.\\
(1)~\textbf{Separation between concepts and implementations:} 
the DCA~\cite{DCA} and its hierarchical version~\cite{HDCA} were motivated by the new advancements in the MPI~3.1 standard, namely MPI  \mbox{one-sided} communication and MPI \mbox{shared-memory}. 
The following question arises:~\textit{Is DCA limited to specific MPI features}?
It is essential to answer this question because only specific MPI runtime libraries fully implement the features of the standard MPI~3.1.
In this manuscript, we separate the idea of DCA and its implementation. 
We highlight specific requirements that a DLS technique needs to fulfill to separate \textit{chunk calculation} that can be distributed across all PEs and the \textit{chunk assignment} that should be synchronized across all PEs.
In contrast to earlier efforts~\cite{DCA,HDCA}, we  introduce and evaluate a \mbox{two-sided} \mbox{MPI-based} implementation of DCA.
This implementation applies to all existing MPI runtime libraries because they fully support \mbox{two-sided} MPI communication.\\
(2)~\textbf{Support for new DLS categories:}
Previously, DCA~\cite{DCA,HDCA} only supported DLS techniques with either \textit{fixed} or \textit{decreasing} chunk size patterns. 
In this extended manuscript, we discuss how DCA supports DLS techniques that calculate \textit{fixed}, \textit{decreasing}, \textit{increasing}, and \textit{irregular} chunk size patterns.\\
(3)~\textbf{DCA in \ourtool{}.} 
We implemented the DCA in an existing \mbox{MPI-based} scheduling library, called \ourtool{}~\cite{LBtool,mohammed2019approach}. 
Initially, all the DLS techniques supported in \ourtool{} were implemented with  a centralized chunk calculation approach~(CCA).
We redesigned and reimplemented  the DLS techniques with DCA in \ourtool{}.
In addition, we added six new DLS techniques and implement them with both CCA and DCA.

The remainder of this work is organized as follows. 
Section~\ref{sec:DLS} contains a review of the selected DLS techniques. 
Existing DLS execution models are reviewed in Section~\ref{sec:models}.
The distributed chunk calculation approach and its execution model are introduced in Section~\ref{sec:DCA}.
We discuss in Section~\ref{sec:DCA} whether the existing mathematical chunk calculation formulas of the selected DLS  techniques support DCA, 
and we show the required mathematical transformations to these chunk calculation formulas to enable DCA.
In Section~\ref{sec:extension}, we present our extensions to \ourtool{} that enable the support of DCA.  
The design of experiments and the experimental results are discussed in Section~\ref{sec:ED}. 
The conclusions and future work directions are outlined in Section~\ref{sec:conclusion}.

%% file: 2.tex
\section{Dynamic Loop Self-scheduling~(DLS)}
\label{sec:DLS}
In scientific applications, loops are the primary source of parallelism~\cite{fang1990dynamic}.
Loop scheduling techniques have been introduced to achieve a balanced load execution of loop iterations. 
When loops have no cross-iteration dependencies, loop scheduling techniques map individual loop iterations to different processing elements aiming to have nearly equal finish times on all processing elements. 
Loop scheduling techniques can be categorized into static and dynamic loop \mbox{self-scheduling}. 
The time when scheduling decisions are taken is the crucial difference between both categories. 
Static loop scheduling (SLS) techniques take scheduling decisions before application execution, while dynamic loop self-scheduling (DLS) techniques take scheduling decisions during application execution.
Therefore, SLS techniques have less scheduling overhead than DLS techniques, and DLS techniques can achieve better load balanced executions than SLS techniques in highly dynamic execution environments.

DLS techniques can further be divided into non-adaptive and adaptive techniques. 
The non-adaptive techniques utilize certain information that is obtained before the application execution. 
The adaptive techniques regularly obtain information during the application execution, and the scheduling decisions are taken based on that new information. 
The adaptive techniques incur a significant scheduling overhead compared to non-adaptive techniques and outperform the non-adaptive ones in highly irregular execution environments.

\begin{figure*}
	\centering
	\includegraphics[clip,trim=0cm 2cm 0cm 3.5cm,  width=\textwidth]{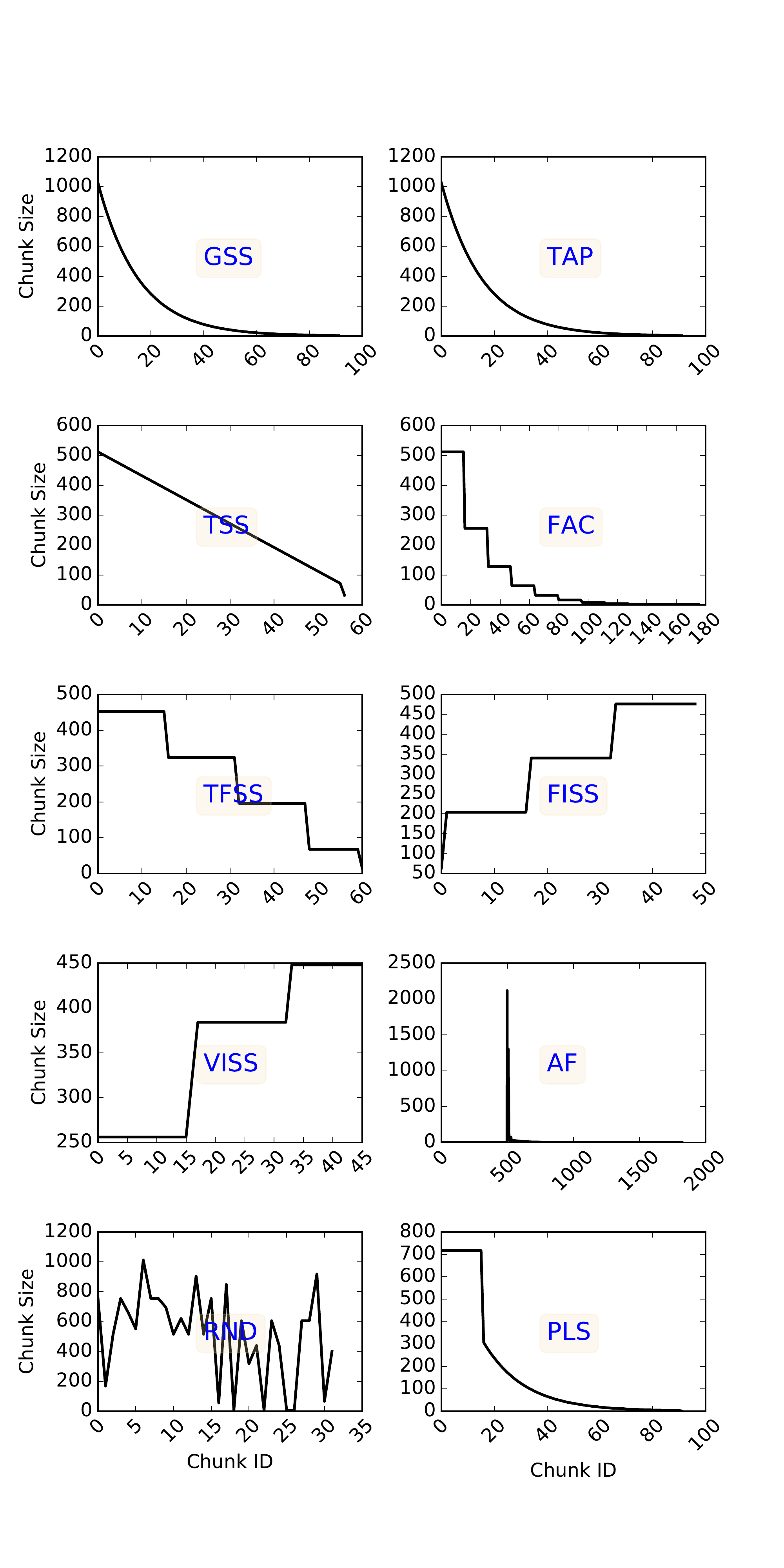}
	\caption{Example of the DLS techniques chunk sizes. The data was obtained
			from the main loop of Mandelbrot~\cite{mandelbrot1980fractal} with 1,000 loop iterations and executing
			on an Intel Xeon processor with 4 MPI ranks. The minimum chunk size is set to be 1 loop iteration.}
	\label{fig:patterns}
\end{figure*}

We consider twelve loop scheduling techniques including static~(STATIC), fixed size chunk~(FSC)~\cite{FSC}, guided self-scheduling~(GSS)~\cite{GSS},
factoring~(FAC)~\cite{FAC}, trapezoid \mbox{self-scheduling}~(TSS)~\cite{TSS},  trapezoid factoring \mbox{self-scheduling}~(TFSS)~\cite{TFSS}, fixed increase \mbox{self-scheduling}~(FISS)~\cite{ISS},
variable increase \mbox{self-scheduling}~(VISS)~\cite{ISS}, tapering~(TAP)~\cite{TAP}, \mbox{performance-based} loop scheduling~(PLS)~\cite{PLS}, and adaptive factoring~(AF)~\cite{AF}.
These techniques employ different strategies to achieve load balanced executions.
As shown in Figure~\ref{fig:patterns}, the calculated chunk sizes may follow fixed, increasing, decreasing , or unpredictable patterns.
Table~\ref{tab:sym} summarizes the notation used in this work to describe how each DLS technique calculates the chunk sizes.

\begin{table}[!h]
	\centering
	\caption{Notation used in the present work}
	\label{tab:sym}
	\resizebox{\textwidth}{!}{
		\begin{tabular}{@{}l|l@{}}
			\textbf{Symbol} & \textbf{Description}                     \\ \hline 
			$N$      	& Total number of loop iterations  \\  
			$P$      	& Total number of processing elements   \\ 
			$S$      	& Total number of scheduling steps  \\  
			$B$      	& Total number of scheduling batches \\ 
			$i$  		& Index of current scheduling step, $0 \leq i \le S-1$\\  
			$b$        	& Index of currently scheduled batch, $0 \leq b \le B-1$\\ 
			$h$			& Scheduling overhead associated with assigning loop iterations\\
			$R_{i}$   	& Remaining loop iterations after \mbox{$i$-th} scheduling step\\  
			$S_{i}$   	& \makecell[l]{Scheduled loop iterations after \mbox{$i$-th} scheduling step \\ $S_{i} + R_{i} = N$} \\  
			$lp_{\text{start}}$ &\makecell[l]{Index of currently executed loop iteration, \\ $0\leq lp_{\text{start}}\leq N-1$}  \\
			$L$ &  \makecell[l]{A DLS technique, \\ $L \in \{STATIC,FSC,GSS, TAP,TSS, FAC,TFSS, FISS,VISS,AF, RND, PLS\}$} \\
			$K^{L}_{0}$	&Size of the largest chunk of a scheduling technique $L$\\  
			$K^{L}_{S-1}$	& Size of the smallest chunk of a scheduling technique $L$ \\  
			$K^{L}_{i}$   	& Chunk size calculated at scheduling step $i$  of a scheduling technique $L$\\  
			$p_{j}$ 	& Processing element  $j$, $0 \leq j \le P-1$\\\  
			$h$ 		& Scheduling overhead for assigning a single iteration\\  
			$\sigma_{p_i}$  	& Standard deviation of the loop iterations' execution times executed on $p_j$\\  
			$\mu_{p_i}$  	& Mean of the loop iterations' execution times executed on $p_j$\\
			$T_{\text{p}}^{\text{loop}}$ 	&  Parallel execution time of the application's parallelized loops  
		\end{tabular}}
	\end{table}

STATIC is a straightforward technique that divides the loop into  $P$ chunks of equal size.
\Eq~\ref{eq:static} shows how STATIC calculates the chunk size. 
Since the scheduling overhead is proportional to the number of calculated chunk sizes,   STATIC incurs the lowest scheduling overhead because it has the minimum number of chunks (only one chunk for each PE).
\begin{equation}
\label{eq:static}
K^{STATIC}_{i} = \frac{N}{P}
\end{equation}

SS~\cite{SS} is a dynamic \mbox{self-scheduling} technique where the chunk size is always one iteration, as shown in~\Eq~\ref{eq:ss}.
SS has the highest scheduling overhead because it has the maximum number of chunks, i.e., the total number of chunks is $N$. 
However, SS can achieve a highly \mbox{load-balanced} execution in highly irregular execution environments.
\begin{equation}
\label{eq:ss}
K^{SS}_{i} = 1
\end{equation}

As a middle point between STATIC and SS,  FSC assumes an optimal chunk size that achieves a balanced execution of loop iterations with the smallest overhead.
To calculate such an optimal chunk size, FSC considers the variability in iterations' execution time and the scheduling overhead of assigning loop iterations to be known before applications' execution.
\Eq~\ref{eq:fsc} shows how FSC calculates the optimal chunk size.
\begin{equation}
\label{eq:fsc}
	K^{FSC}_{i} = \frac{\sqrt{2}\cdot N \cdot h}{\sigma \cdot P \cdot  \sqrt{\log{P}}}
\end{equation}

GSS~\cite{GSS} is also a compromise between the highest load balancing that can be achieved using SS and the lowest scheduling overhead incurred by STATIC.
Unlike FSC, GSS assigns decreasing chunk sizes to balance loop executions among all PEs.
At every scheduling step, GSS assigns a chunk that is equal to  the number of remaining loop iterations divided by the total number of PEs, as shown in~\Eq~\ref{eq:gss}.
\begin{equation}
\label{eq:gss}
\begin{split}
&K^{GSS}_{i} = \frac{R_i}{P}\text{, where} \\
&R_i=  N - \sum_{j=0}^{i-1} k^{GSS}_j
\end{split}
\end{equation}

TAP~\cite{TAP} is based on a probabilistic analysis that represents a general case of GSS.
It considers the average of loop iterations execution time~$\mu$ and their standard deviation~$\sigma$ to achieve a higher load balance than GSS. 
\Eq~\ref{eq:tap} shows how TAP tunes the GSS chunk size based on ~$\mu$ and~$\sigma$.
\begin{equation}
\label{eq:tap}
\begin{split}
&K^{TAP}_{i} = K^{GSS}_i + \frac{v^2_{\alpha}}{2} - v_\alpha \cdot \sqrt{2 \cdot K^{GSS}_i + \frac{v^2_\alpha}{4}}\text{, where}\\
&v_\alpha =  \frac{\alpha \cdot \sigma}{\mu}
\end{split}
\end{equation}

TSS~\cite{TSS} assigns decreasing chunk sizes similar to GSS. 
However, TSS uses a linear function to decrement chunk sizes. 
This linearity results in low scheduling overhead in each scheduling step compared to GSS.
\Eq~\ref{eq:tss} shows the linear function of TSS.
\begin{equation}
\label{eq:tss}
\begin{split}
&K^{TSS}_{i}= K^{TSS}_{i-1} - \floor*{ \frac{K^{TSS}_0-K^{TSS}_{S -1}}{S-1} }\text{, where}\\
&S = \ceil*{\frac{2\cdot N}{K^{TSS}_0+K^{TSS}_{S-1}}}\\ 
&K^{TSS}_0=\ceil*{ \frac{N}{2\cdot P} }\-\-, K^{TSS}_{S-1}=1
	\end{split}
\end{equation}

FAC~\cite{FAC} schedules the loop iterations in batches of equally-sized chunks. 
FAC evolved from comprehensive probabilistic analyses and it assumes prior knowledge about~$\mu$ and~$\sigma$ their mean execution time. 
Another practical implementation of FAC denoted FAC2, assigns half of the remaining loop iterations for every batch, as shown in \Eq~\ref{eq:fac}. 
The initial chunk size of FAC2 is half of the initial chunk size of GSS. 
If more \mbox{time-consuming} loop iterations are at the beginning of the loop, FAC2 may better balance their execution than GSS.
\begin{equation}
\label{eq:fac}
	\begin{split}
	&K^{FAC2}_i = \left\{ \begin{array}{ll} \ceil*{ \frac{R_{i}}{2 \cdot P}},  \text{if} \ i\mod P = 0 \\ K^{FAC2}_{i-1},  \text{otherwise.} \end{array} \right. \text{, where}\\
	&R_i=  N - \sum_{j=0}^{i-1} k^{FAC2}_j
	\end{split}
\end{equation}

TFSS~\cite{TFSS}  combines certain characteristics of TSS~\cite{TSS} and FAC~\cite{FAC}.
Similar to FAC, TFSS schedules the loop iterations in batches of \mbox{equally-sized} chunks.
However, it does not follow the analysis of FAC, i.e.,  every batch is not  half of the remaining number of iterations.
Batches in TFSS decrease linearly, similar to chunk sizes in TSS.
As shown in \Eq~\ref{eq:tfss}, TFSS calculates the chunk size as the sum of the next P chunks that would have been computed by the TSS divided by P.
\begin{equation}
\label{eq:tfss}
K^{TFSS}_i=\left\{ \begin{array}{ll}  \frac{\sum_{j=i}^{i+P} K^{TSS}_{j-1} }{P} & \text{if} \ i\-\-\mod P = 0 \\ K^{TFSS}_{i-1}, & \text{otherwise.} \end{array} \right.
\end{equation}

GSS~\cite{GSS}, TAP~\cite{TAP}, TSS~\cite{TSS}, FAC~\cite{FAC}, and TFSS\cite{TFSS} employ a  decreasing chunk size pattern.
This pattern introduces additional scheduling overhead due to the small chunk sizes towards the end of the loop execution. 
On \mbox{distributed-memory} systems, the additional scheduling overhead is more substantial than on \mbox{shared-memory} systems.
Fixed increase size chunk~(FISS)~\cite{ISS} is the first scheduling technique devised explicitly for \mbox{distributed-memory} systems.
FISS follows an increasing chunk size pattern calculated as in \Eq~\ref{eq:fiss}. 
FISS depends on an initial value $B$ defined by the user (suggested to be equal to the total number of batches of FAC).
\begin{equation}
\label{eq:fiss}
\begin{split}
&K^{FISS}_i= K^{FISS}_{i-1} + \ceil{ \frac{2 \cdot N \cdot (1 - \frac{B}{2 + B}) }{P \cdot B \cdot (B-1)}}\text{, where}\\
&K^{FISS}_0 = \frac{N}{(2+B) \cdot P}
\end{split}
\end{equation}

VISS~\cite{ISS} follows an increasing pattern of chunk sizes. 
Unlike FISS, VISS relaxes the requirement of defining an initial value $B$.
VISS works similarly to FAC2, but instead of decreasing the chunk size, VISS  increments the chunk size by a factor of two per scheduling step.
\Eq~\ref{eq:viss} shows the chunk calculation of VISS.
\begin{equation}
\begin{split}
\label{eq:viss}
&K^{VISS}_i = \left\{ \begin{array}{ll} K^{VISS}_{i-1} + \frac{K^{VISS}_{i-1} }{2} & \text{if} \ i \mod P = 0 \\ K^{VISS}_{i-1}, & \text{otherwise.} \end{array} \right.\text{, where}\\
&K^{VISS}_0 = 	K^{FISS}_0  
\end{split}
\end{equation}

AF~\cite{AF} is an adaptive DLS technique based on FAC.
However, in contrast to FAC, AF learns both $\mu$ and $\sigma$ for each computing resource during application execution to ensure full adaptivity to all factors that cause load imbalance.
AF does not follow a specific pattern of chunk sizes. 
AF adapts chunk size based on the continuous updates of $\mu$ and $\sigma$ during applications execution. 
Therefore, the pattern of AF's chunk sizes is unpredictable.
\Eq~\ref{eq:af} shows the chunk calculation of AF.
\begin{equation}
\label{eq:af}
\begin{split}
&K^{AF}_i = \frac{ D + 2 \cdot E \cdot R_i - \sqrt{D^2 + 4 \cdot D \cdot E \cdot R_i}   }{2 \mu_{p_i}}\text{,where}\\
&D = \sum_{p_i=1}^{P} \frac{\sigma^2_{p_i}}{\mu_{p_i}}\\ 
&E = \left(\sum_{p_i=1}^{P} \frac{1}{\mu_{p_i}}\right)^{-1}
\end{split}
\end{equation}

RND~\cite{RND} is a DLS technique that utilizes a uniform random distribution to arbitrarily choose a chunk size between specific lower and upper bounds. 
The lower and the upper bounds were suggested to be $\frac{N}{100 \cdot P }$ and $\frac{N}{2 \cdot P}$, respectively~\cite{RND}.
In the current work, we suggest a lower and an upper bound as $1$ and $\frac{N}{P}$, respectively.
These bounds make RND have an equal probability of selecting any chunk size between the chunk size of STATIC and the chunk size of SS, which are the two extremes of DLS techniques in terms of scheduling overhead and load balancing.
\Eq~\ref{eq:rnd} represents the integer range of the RND chunk sizes.
\begin{equation}
\label{eq:rnd}
K^{RND}_i \in \bigl[ 1,N/P\bigr]
\end{equation}

PLS~\cite{PLS} combines the advantages of SLS and DLS.
It divides the loop into two parts. 
The first loop part is scheduled statically.
In contrast,  the second part is scheduled dynamically using GSS.
The static workload ratio~(SWR) is used to determine the amount of the iterations to be statically scheduled.
SWR~is calculated as the ratio between minimum and maximum iteration execution time of five randomly chosen iterations. 
PLS also uses a performance function~(PF) to statically assign parts of the workload to each processing element~$p_j$ based on the PE's speed and its current CPU load. 
In the present work, all PEs are assumed to have the same load during the execution. 
This assumption is valid given the exclusive access to the HPC infrastructure used in this work. 
\Eq~\ref{eq:pls} shows the chunk calculation of PLS.
\begin{equation}
\label{eq:pls}
\begin{split}
&K^{PLS}_i = \left\{ \begin{array}{ll} \frac{N \cdot SWR}{P}, & \text{if} \ R_i \textgreater N- (N \cdot SWR)  \\ K^{GSS}_{i}, & \text{otherwise.} \end{array} \right.\text{, where}\\
&SWR= \frac{\text{minimum iteration execution time}}{ \text{maximum iteration execution time}}
\end{split}
\end{equation}

Table~\ref{tab:chunks} shows the chunk sizes generated by each technique.
We obtain these chunks by assuming that the total number of iterations $N$ is 1,000  and the total number of PEs $P$ is 4.
In addition to these two parameters, we consider other parameters required by each DLS technique.
For instance, FSC requires the scheduling overhead $h$, which is considered to be 0.013716 seconds. 
TAP requires $\mu$, $\sigma$, and $\alpha$ that are assumed to be 0.1, 0.0005,  and 0.0605 seconds, respectively.
For FISS and VISS, we consider $B$ and  $X$  to be 3 and 4.
For PLS, we assume the $SWR$ ratio be 0.7.
\begin{table}[]
	\centering

	\caption{Chunk sizes for the selected DLS techniques considered in the current work for the main loop of Mandelbrot~\cite{mandelbrot1980fractal} with 1,000 loop iterations and executing
			on an Intel Xeon processor with 4 MPI ranks.}
	\label{tab:chunks}
	\resizebox{\textwidth}{!}{
		\begin{tabular}{l|l|l}
			\textbf{Technique} & \textbf{Chunk sizes} & \begin{tabular}[c]{@{}l@{}}\textbf{Total number of}\\ \textbf{chunks}\end{tabular} \\ \hline
			STATIC & 250, 250, 250, 250 & 4 \\ \hline
			SS & 1, 1, 1, $\cdots$, 1 & 1000 \\ \hline
			FSC & 17, 17, 17, 	$\cdots$, 14 & 59 \\ \hline
			GSS & \begin{tabular}[c]{@{}l@{}}250, 188,  141, 106, 80, 60, 45, 34, 26, 19, 15, 11, 8, 6, 5, 4, 2\end{tabular} & 17 \\ \hline
			TAP & \begin{tabular}[c]{@{}l@{}}250, 188,  141, 106, 80, 60, 45, 34,  26, 19, 15, 11, 8, 6, 5, 3, 3\end{tabular} & 17 \\ \hline
			TSS & \begin{tabular}[c]{@{}l@{}}125, 117, 109, 101, 93, 85, 77, 69, 61, 53, 45, 37, 28\end{tabular} & 13 \\ \hline
			FAC & \begin{tabular}[c]{@{}l@{}}125, 125, 125, 125, 63, 63, 63, 63, 32, 32, 32, 32, 16, 16, 16, 16, \\8, 8, 8, 8, 4, 4, 4, 4, 2, 2, 2, 2\end{tabular} & 28 \\ \hline
			TFSS & 113, 113, 113, 113,  81, 81, 81, 81, 49, 49, 49, 49, 17, 11 & 14 \\ \hline
			FISS & 50, 50, 50, 50, 83, 83, 83, 83, 116, 116, 116, 116, 4 & 13 \\ \hline
			VISS & 62, 62, 62, 62, 93, 93, 93, 93, 108, 108, 108, 56 & 12 \\ \hline
			AF & \begin{tabular}[c]{@{}l@{}}1,1, $\cdots$, 3544, 3544, 2410, 1785, 235, 202, 179, 321, 247, \\ 267, 197, 222, 202, 182, 157, 157, 144, 128, 126, 116, 105, 102, \\ 86, 90, 89, 78, 72, 69, 65, 61, 57, 53, 50, 49, 45, 42, 40, 38, 36,\\  37, 33, 33, 29, 28, 28, 24, 23, 22, 21, 21, 19, 18, 17, 16, 16, 15,\\  14, 13, 12, 13, 12, 11, 11, 10, 10, 9, 9, 8, 8, 7, 7, 7, 6, 6, 5, 6, 5, \\ 5, 5, 5, 4, 4, 4, 4, 4, 4, 3, 3, 3, 3, 3, 3, 3, 3, 2, 2, 2, 2, 2, 2, 2, 2, 2,\\ 1, 1, $\cdots$, 1 \end{tabular} & 316 \\ \hline
			RND & \begin{tabular}[c]{@{}l@{}}17, 84, 16, 36, 220, 64, 45, 81, 56, 210, 34, 29,  8, 100\end{tabular} & 14 \\ \hline
			PLS & \begin{tabular}[c]{@{}l@{}}175, 175, 175, 175, 75, 57, 43, 32, 24, 18, 14, 11, 8, 6, 5, 4, 3\end{tabular} & 17 \\ \hline
		\end{tabular}
		
	}
\end{table}

%% file: 3.tex
\newpage
\section{DLS Implementation Approaches}
\label{sec:models}
In the current work, the \mbox{self-scheduling} aspect of the DLS techniques means that once a PE becomes free, it \textbf{calculates} a new chunk of loop iterations to be executed.
The calculated chunk size is not associated with a specific set of loop iterations.
Since the DLS techniques assume a central work queue, 
the PE must synchronize with all other PEs to \textbf{\mbox{self-assign}} unscheduled loop iterations. 
We can conclude that there are two operations at every scheduling step: (1) \textit{chunk calculation} and (2) \textit{chunk assignment}.
In principle, only the \textit{chunk assignment} requires a sort of global synchronization between all PEs, while the chunk calculation does not require synchronization and can be distributed across all PEs.
In practice, existing DLS implementation approaches, especially for \mbox{distributed-memory} systems,  do not consider the separation between chunk calculation and chunk assignment. 
Hence, the \mbox{master-worker} execution model dominates all existing DLS implementation approaches.

\begin{figure*}
	\captionsetup[subfigure]{justification=centering}
	\centering
	\begin{subfigure}{0.6\textwidth}
		\centering
		\includegraphics[clip,trim=0cm 0cm 0cm 0cm, width=0.9\textwidth]{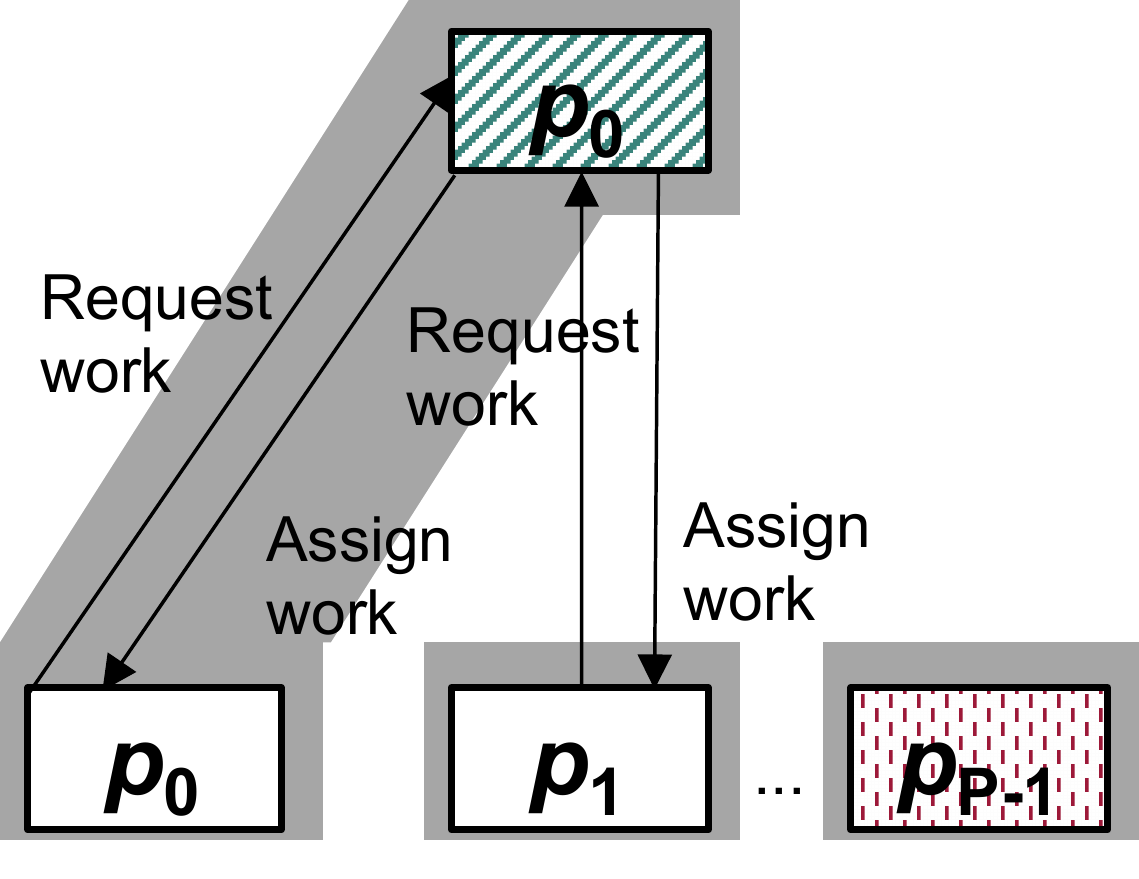}
		\subcaption{Conventional master-worker execution model}
		\label{fig:con-master-worker}%
	\end{subfigure}%
	
	\begin{subfigure}{0.6\textwidth}
		\centering
		\includegraphics[width=0.9\textwidth]{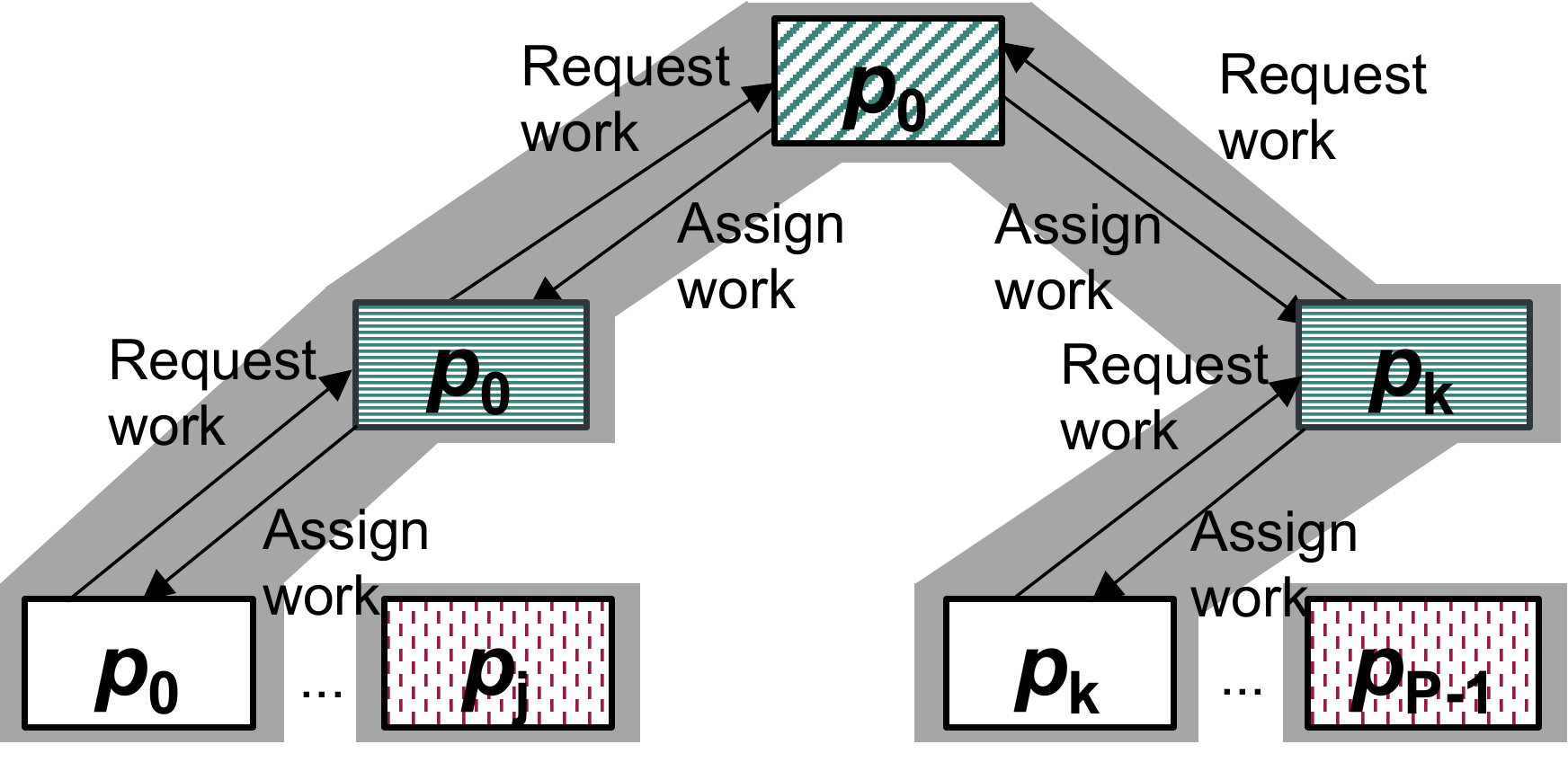}
		\subcaption{Global and local masters are located on a single physical compute node}
		\label{fig:con-h-master-same}%
	\end{subfigure}%
	
	\begin{subfigure}{0.6\textwidth}
		\centering
		\includegraphics[width=0.9\textwidth]{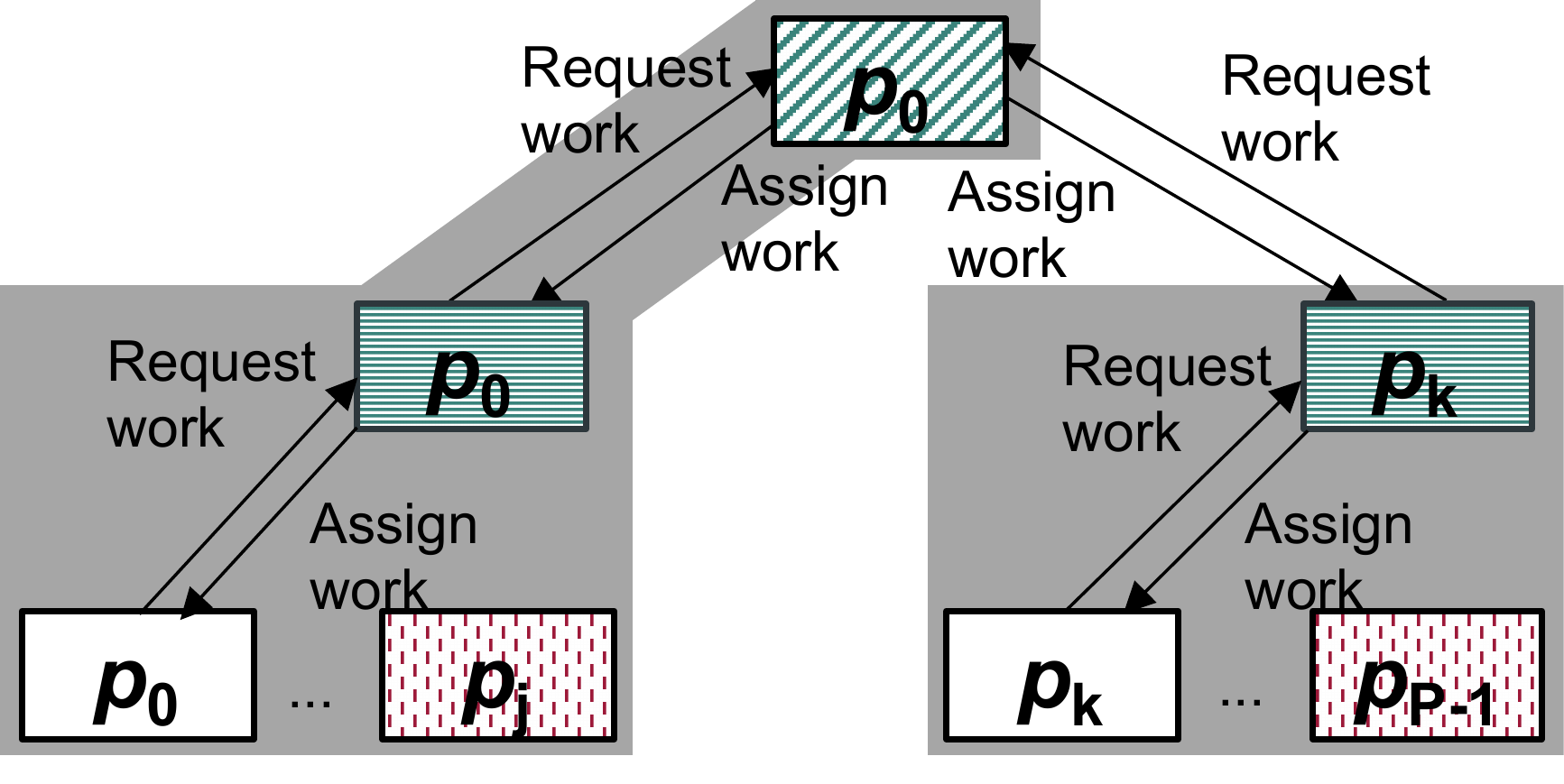}
		\subcaption{Local masters are distributed across multiple physical compute nodes}
		\label{fig:con-h-master-dis}%
	\end{subfigure}
	
	\begin{subfigure}{\textwidth}
		\centering
		\includegraphics[clip, trim=0cm 0cm 0cm 0cm,width=\textwidth]{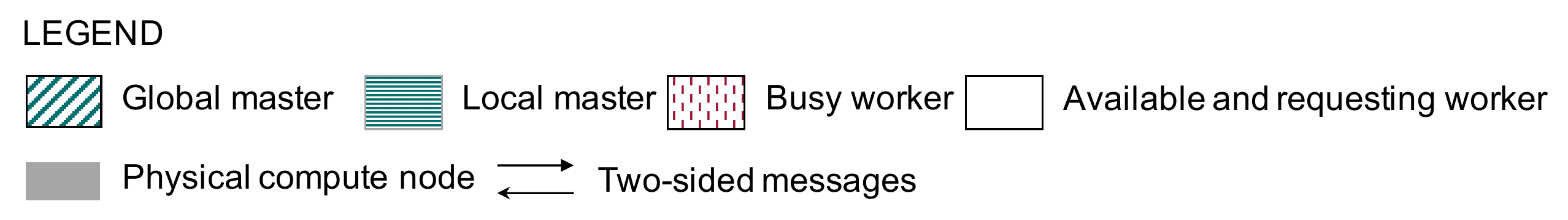}
	\end{subfigure}%
	
	\caption{Variants of the \mbox{master-worker} execution model as reported in the literature. Replication of certain processing elements is just to indicate their double role where the master participates in the computation as a worker.}
	\label{fig:models}
\end{figure*}

The distributed \mbox{self-scheduling} scheme~(DSS)~\cite{TFSS} is an example of employing the master-worker model to implement DLS techniques for \mbox{distributed-memory} systems. 
DSS relies on the \mbox{master-worker} execution model,  similar to the one illustrated in Figure~\ref{fig:con-master-worker}.
DSS enables the master to consider the speed of the processing elements and their loads when assigning new chunks.
DSS was later enhanced by a hierarchical distributed self-scheduling scheme (HDSS)~\cite{DHTSS} that employs a hierarchical \mbox{master-worker} model, as illustrated in Figure~\ref{fig:con-h-master-same}.
DSS and HDSS assume a dedicated master configuration in which the master PE is reserved for handling the worker requests.
Such a configuration may enhance the scalability of the proposed \mbox{self-scheduling} schemes.
However, it results in low CPU utilization of the master.
HDSS~\cite{DHTSS} suggested deploying the \mbox{global-master} and the \mbox{local-master} on one physical computing node with multiple processing elements to overcome the low CPU utilization of the master~(see Figure~\ref{fig:con-h-master-same}). 
DSS and HDSS were implemented using MPI \mbox{two-sided} communications.
In both DSS and HDSS, the master is a central entity that performs both the \textit{chunk calculation} and the \textit{chunk assignment}.

Another \mbox{MPI-based} library that implements several DLS techniques is called the \textit{load balancing tool}~(LB tool)~\cite{LBtool}.
At the conceptual level, the LB tool is based on a \mbox{single-level} \mbox{master-worker} execution model~(see Figure~\ref{fig:con-master-worker}).   
However, it does not assume a dedicated master. 
It introduces the \textit{breakAfter} parameter, which is \mbox{user-defined}, and indicates how many iterations the master should execute before serving pending worker requests.
This parameter is required for dividing the time of the master  between computation and servicing of worker requests. 
The optimal value of this parameter is application- and \mbox{system-dependent}.
The LB tool also employs \mbox{two-sided} MPI communications. 

\ourtool{}~\cite{mohammed2019approach,mohammed2020simas} is an extension of the LB tool~\cite{LBtool} that includes certain bug fixes and additional DLS techniques.
Both LB and \ourtool{} employ a \mbox{master-worker} execution in which the master is a central entity that performs both of chunk calculation and the chunk assignment operations.

The  dynamic load balancing library~(DLBL)~\cite{DLBLtool} is another \mbox{MPI-based} library used for cluster computing.
It is based on a parallel runtime environment for multicomputer applications~(PREMA)~\cite{PREMA}. 
DLBL is the first tool that employed MPI \mbox{one-sided} communication for implementing DLS techniques. 
Similar to the LB tool, the DLBL employs a \mbox{master-worker} execution model.
The master expects work requests. 
It then calculates the size of the chunk to be assigned and, subsequently, calls a handler function on the worker side.
The worker is responsible for obtaining the new chunk data without any further involvement from the master.
This means that the master is still a central entity that performs both of chunk calculation and chunk assignment. 

\begin{figure}[!h]
	\centering
	\includegraphics[clip, trim=0cm 1.2cm 0.1cm 0cm,width=\columnwidth]{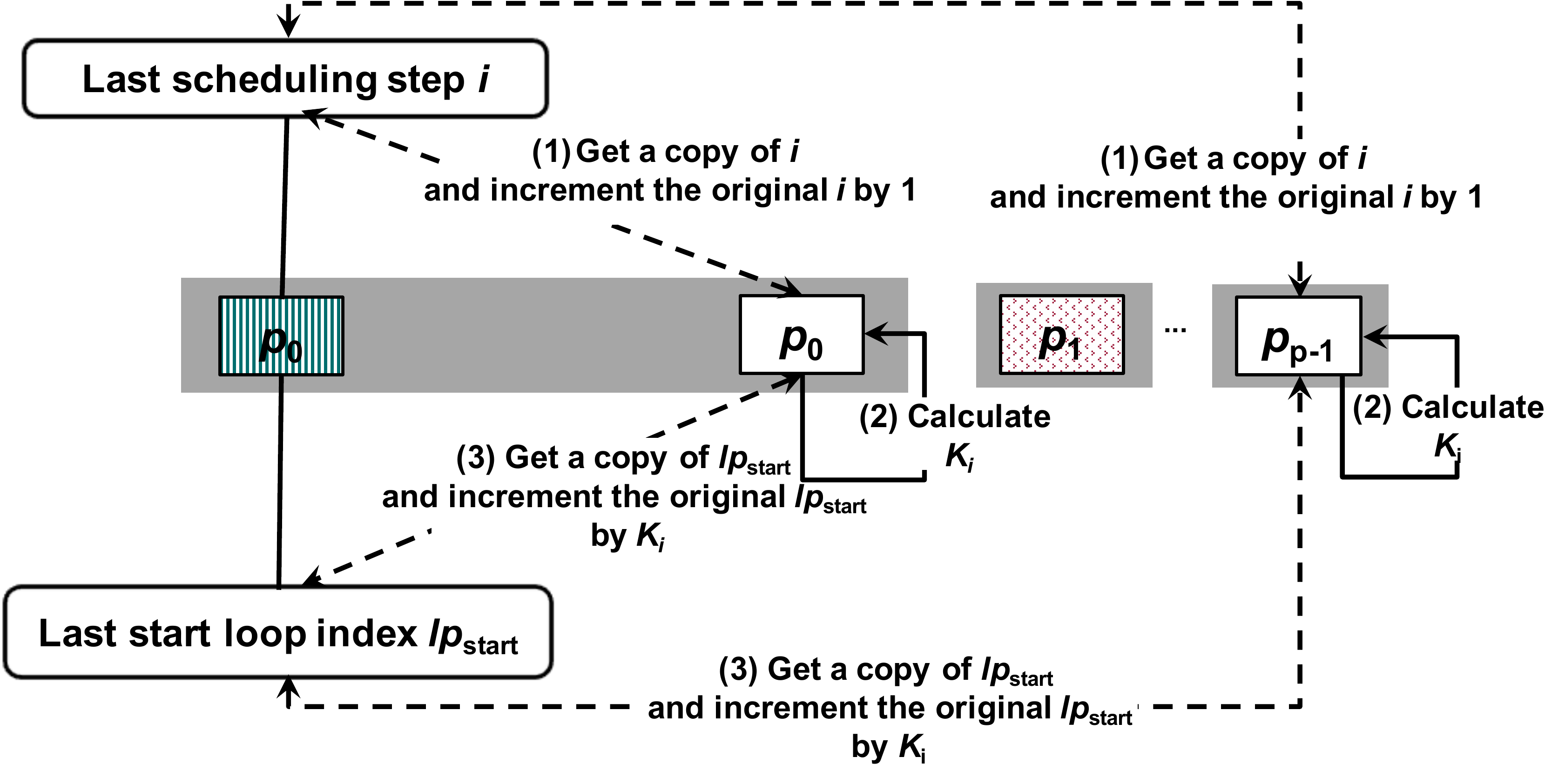}
	\includegraphics[width=\columnwidth]{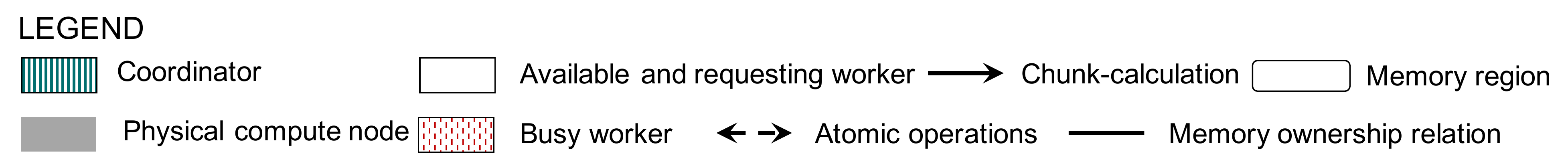}
	\caption{The distributed chunk calculation approach~(DCA) using MPI RMA and \mbox{passive-target} synchronization. }
	\label{fig:DCA}
	\end{figure}

The latest advancements in the MPI~3.1 standard, namely the revised and the clear semantics of the MPI RMA (\mbox{one-sided} communication)~\cite{RMAhoefler,zhao2016scalability}, enabled its usage in different scientific applications~\cite{hammond2014implementing,shan2017experiences,zhou2016asynchronous}.  
This motivated our earlier work~\cite{DCA} that introduced DCA.
The DCA  does not require the \mbox{master-worker} execution scheme~\cite{DCA}.
Using MPI RMA, DCA makes one processing element, called \textit{coordinator}, store global scheduling information such as the index of the latest scheduling step $i$ and the index of the previously scheduled loop iteration $lp_{start}$.
The coordinator entity shares the memory address space where the global scheduling information is stored with all workers.
Figure~\ref{fig:DCA} shows that with certain exclusive load and store operations to the shared memory address space, all entities can simultaneously calculate and assign themselves chunks of non-overlapping loop iterations.
The following question arises:~\textit{Is DCA limited to specific MPI features}?
It is essential to answer this question because only specific MPI runtime libraries fully implement the features of the MPI~3.1 standard.

%% file: 4.tex
\section{Distributed Chunk Calculation Approach (DCA)}
\label{sec:DCA}

The idea of DCA is to ensure that the calculated chunk size at a specific PE does not rely on any information about the chunk size  calculated at any other PE.
The chunk calculation formulas (\Eq~\ref{eq:static} to~\ref{eq:pls})  can be classified into straightforward and recursive.
\textbf{A straightforward chunk calculation formula} only requires some constants and input parameters.
\textbf{A recursive chunk calculation formula} requires information about previously calculated chunk sizes.  
For instance, STATIC, SS, FSC, and RND have  straightforward chunk calculation formulas that do not require any information about previously calculated chunks, while
GSS~\cite{GSS}, TAP~\cite{TAP}, TSS~\cite{TSS}, FAC~\cite{FAC}, TFSS~\cite{TFSS}, FISS~\cite{ISS}, VISS~\cite{ISS}, AF~\cite{AF}, and PLS~\cite{PLS} employ recursive chunk calculation formulas. 
Certain transformations have been required to convert these recursive formulas into straightforward formulas to enable DCA.
For GSS and FAC, the transformations were already introduced in the literature~\cite{FAC} (\Eq~\ref{eq:gss2} and~\ref{eq:fac2}).
\begin{align}
\label{eq:gss2}
&K^{\prime^{GSS}}_{i} = \left \lceil \left(\frac{P-1}{P}\right)^i \cdot \frac{N}{P} \right \rceil \\
\label{eq:fac2}
&K^{\prime^{FAC2}}_{i} = \left \lceil \left(\frac{1}{2}\right)^{i_{\text{new}}} \cdot \frac{N}{P} \right \rceil, \ \ i_{\text{new}}= \left \lfloor \frac{i}{P} \right \rfloor+1
\end{align}

As shown in \Eq~\ref{eq:tap}, TAP calculates $K^{GSS}_i$ and tunes that value based on $\mu$, $\sigma$, and $\alpha$.
Based on \Eq~\ref{eq:gss2}, the chunk calculation formula of TSS can be expressed as a straightforward formula as follows.
\begin{equation}
\label{eq:tap2}
\begin{split}
&K^{\prime^{TAP}}_{i}  = K^{\prime^{GSS}}_{i} + \frac{v^2_{\alpha}}{2} - v_\alpha \cdot \sqrt{2 \cdot K^{\prime^{GSS}}_{i} + \frac{v^2_\alpha}{4}}\text{, where}\\
&v_\alpha =  \frac{\alpha \cdot \sigma}{\mu}
\end{split}
\end{equation}

For TSS, a straightforward formula for the chunk calculation is shown in \Eq~\ref{eq:tss2}.
\begin{equation}
\label{eq:tss2}
K^{\prime^{TSS}}_{i} = K^{TSS}_{0} - i \cdot \floor{ \frac{K^{TSS}_0-K^{TSS}_{S -1}}{S-1}} 
\end{equation}
The mathematical derivation that converts \Eq~\ref{eq:tss} into \Eq~\ref{eq:tss2} is as follows.
The TSS chunk calculation formula can be represented as follows, where C is a constant.
\begin{equation*}
\begin{split}
&K^{TSS}_{i}= K^{TSS}_{i-1} - C\\
&C=\floor*{ \frac{K^{TSS}_0-K^{TSS}_{S -1}}{S-1} }\\
&K^{TSS}_{1} = K^{TSS}_{0} - C \\
&K^{TSS}_{2}= K^{TSS}_{1} - C = (K^{TSS}_{0} - C) - C =  K^{TSS}_{0} - 2 \cdot C  \\
&K^{TSS}_{i} = K^{TSS}_{0} - i \cdot C \\
&K^{TSS}_{i} = K^{TSS}_{0} - i \cdot \floor{ \frac{K^{TSS}_0-K^{TSS}_{S -1}}{S-1}} = K^{\prime^{TSS}}_{i}
\end{split}
\end{equation*}

TFSS~\cite{TFSS} is devised based on TSS~\cite{TSS} and FAC~\cite{FAC}.
Therefore, the straightforward formula of TSS (see \Eq~\ref{eq:tss}) can be used to derive the straightforward formula of TFSS, as shown in \Eq~\ref{eq:tfss2}.
\begin{equation}
\label{eq:tfss2}
K^{\prime^{TFSS}}_{i} = \frac{\sum_{j=i}^{i+P} K^{\prime^{TSS}}_{j-1} }{P}
\end{equation}

For FISS~\cite{ISS}, a straightforward formula for the chunk calculation is shown in \Eq~\ref{eq:fiss2}.
\begin{equation}
\label{eq:fiss2}
K^{\prime^{FISS}}_{i}= K^{FISS}_{0} + i \cdot \ceil{ \frac{2 \cdot N \cdot (1 - \frac{B}{2 + B}) }{P \cdot B \cdot (B-1)}}
\end{equation}
The mathematical derivation  that converts \Eq~\ref{eq:fiss} into \Eq~\ref{eq:fiss2} is as follows.
Given that $A$ is a constant, the FISS chunk calculation formula can be represented as follows, where $C$ is a constant.
\begin{equation*}
\begin{split}
&K^{FISS}_i= K^{FISS}_{i-1} + C\\
&C=\ceil{ \frac{2 \cdot N \cdot (1 - \frac{B}{2 + B}) }{P \cdot B \cdot (B-1)}}\\
&K^{FISS}_{1} = K^{FISS}_{0} + C \\
&K^{FISS}_{2}= K^{FISS}_{1} + C = (K^{FISS}_{0} + C) + C =  K^{FISS}_{0} + 2 \cdot C  \\
&K^{FISS}_{i} = K^{FISS}_{0} + i \cdot C \\
&K^{FISS}_{i} = K^{FISS}_{0} + i \cdot \ceil{ \frac{2 \cdot N \cdot (1 - \frac{B}{2 + B}) }{P \cdot B \cdot (B-1)}} = K^{\prime^{FISS}}_{i}
\end{split}
\end{equation*}

For VISS~\cite{ISS}, a straightforward formula for the chunk calculation is shown in \Eq~\ref{eq:viss2}.
\begin{equation}
\begin{split}
\label{eq:viss2}
&K^{\prime^{VISS}}_i = K^{FISS}_0 \cdot \frac{1-(0.5)^{i_{new}}}{0.5}, \text{where} \ \  i \textgreater 0\\
& i_{new} = i \mod P\\ 
&K^{\prime^{VISS}}_0 = K^{FISS}_{0}
\end{split}
\end{equation}
To derive \Eq~\ref{eq:viss2}, we calculate $K^{VISS}_1$, $K^{VISS}_2$, and  $K^{VISS}_3$, according \Eq~\ref{eq:viss}.
\begin{equation*}
\begin{split}
&K^{VISS}_1 = K^{FISS}_0 + \frac{K^{FISS}_0}{2}, \text{assume} \ \  K^{FISS}_0 = a  \\
&K^{VISS}_1 = a + \frac{a}{2}\\
&K^{VISS}_2 = K^{VISS}_1 + \frac{K^{VISS}_1}{2}= (a + \frac{a}{2}) + (\frac{a + \frac{a}{2}}{2})\\
&K^{VISS}_3 = K^{VISS}_2 + \frac{K^{VISS}_2}{2}= ((a + \frac{a}{2}) + (\frac{a + \frac{a}{2}}{2})) + \frac{((a + \frac{a}{2}) + (\frac{a + \frac{a}{2}}{2}))}{2}\\
&\text{According to the geomertic summation theorem}\\
&K^{VISS}_i = K^{FISS}_0  \cdot \frac{1 - (0.5)^i}{0.5}\\
&\text{since VISS assigns chunks in batches}\\
&K^{VISS}_i =  K^{FISS}_0 \cdot \frac{1-(0.5)^{i_{new}}}{0.5} = K^{\prime^{VISS}}_i , \text{where} \  i \textgreater 0, \\
&\text{and } i_{new} = i  \mod P.\\ 
\end{split}
\end{equation*}

For PLS, the loop iteration space is divided into two parts. 
In the first part,  the PLS chunk calculation formula is equivalent to STATIC, i.e., the chunk calculation formula is a straightforward formula that is ready to support DCA. 
In the second part, PLS uses the GSS chunk calculation formula.
Therefore, we  replace $K_i^{GSS}$ in \Eq~\ref{eq:pls} with $K^{\prime^{GSS}}_{i}$ from \Eq~\ref{eq:gss2} to derive the PLS chunk calculation (\Eq~\ref{eq:pls2}).
\begin{equation}
\label{eq:pls2}
\begin{split}
&K^{\prime^{PLS}}_i = \left\{ \begin{array}{ll} \frac{N \cdot SWR}{P}, & \text{if} \ R_i \textgreater N- (N \cdot SWR)  \\ K^{\prime^{GSS}}_{i}, & \text{otherwise.} \end{array} \right.\\
\end{split}
\end{equation}

AF adapts the calculated chunk size according $\mu_{p_i}$ and $\sigma_{p_i}$, which can be determined only during loop execution.
Moreover, at every scheduling step, AF uses $R_i$ with  $\mu_{p_i}$ and $\sigma_{p_i}$ to calculate the chunk size.
This leads to an unpredictable pattern of chunk sizes and makes it impossible to find a straightforward formula for AF. 
Accordingly, we could not determine a way to implement AF with a \textit{fully distributed chunk calculation}.
In our implementation, AF with DCA requires additional synchronization of $R_i$ across all PEs. 
All PEs can simultaneously calculate $D$ and $E$ from \Eq~\ref{eq:af}. 
However, each PE needs to synchronize with all other PEs to calculate each $K_{i}^{AF}$.

%% file: 5.tex
\section{DCA implementation into \ourtool{}}
\label{sec:extension}
\ourtool{}~\footnote{\url{https://github.com/unibas-dmi-hpc/DLS4LB.git}}~\cite{mohammed2019approach,mohammed2020simas} is a recent \mbox{MPI-based} library for loop scheduling and dynamic load balancing.
\ourtool{} extends the LB tool~\cite{LBtool} by including certain bug fixes and additional DLS techniques.
\ourtool{} has been used to enhance the performance of various scientific applications~\cite{mohammed2020}.
In this work, we extend the \ourtool{} in two directions: (1)~We enable the support of DCA. 
All the DLS techniques originally supported in \ourtool{} were implemented with  a centralized chunk calculation approach~(CCA).
We redesign and reimplement them with DCA.
(2)~We add six additional DLS techniques and implement them with CCA and DCA.

While \ourtool{} schedules independent loop iterations across multiple MPI processes, it assumes that each MPI process has access to the data associated with the loop iterations it executes. 
The simplest way to ensure the validity of that assumption is to replicate the data of all loop iterations across all MPI processes.
Users can also centralize or distribute  data of the loop iterations across all MPI processes.
In this case, however,  users need to provide a way to their application to exchange the required data associated with the loop iterations.

The  \ourtool{} has six API functions: \texttt{DLS\_Parameters\_Setup}, \texttt{DLS\_StartLoop},
\texttt{DLS\_Terminated}, \texttt{DLS\_StartChunk}, \texttt{DLS\_EndChunk}, and \texttt{DLS\_EndLoop}.
One can use these API functions as in Listing~\ref{algo:ourtool}. 
\begin{algorithm}[!h]
	\#include$\textless$mpi.h$\textgreater$;\newline
	\#include$\textless$\ourtool.h$\textgreater$\;
	\textbf{main()}\{\\
	\setcounter{AlgoLine}{0}
	\ShowLn	
	\small{\texttt{.../*application code*/}}\\
	\ShowLn	
	int mode = DECENTRALIZED; \small{\texttt{/*or CENTRALIZED*/}}\\
	\ShowLn	
	Configure\_Chunk\_Calculation\_Mode($mode$)\;
	\ShowLn	
	DLS\_Parameters\_Setup($params$); \small{\texttt{/*includes number of tasks, scheduling method, scheduling parameters $\mu$, $\sigma$, $\cdots$ etc*/}}\\
	\ShowLn	
	DLS\_StartLoop( info, start\_index,end\_index, scheduling\_method )\;
	\While{(!DLS\_Terminated($info$))}
	{
		\ShowLn
		DLS\_StartChunk($info$, $lp_{start}$, $chunk\_size$ )\; 
		\ShowLn
		\small{\texttt{/*application code to process loop from $lp_{start}$ to $lp_{start} + chunk\_size$ */}}\\
		\ShowLn
		 DLS\_EndChunk($info$)\; 
	}
	\ShowLn
	DLS\_EndLoop($info$, $scheduled\_tasks$, $total\_time$);
	\}
	\caption{Usage of \ourtool{} for loop scheduling  and dynamic load balancing in scientific applications}
	\label{algo:ourtool}
\end{algorithm}
For backward compatibility reasons, our extension of \ourtool{} maintained these six APIs. 
However, we added a new API: \texttt{Configure\_Chunk\_Calculation\_Mode} that selects between CCA and DCA.
We changed the functionality of each of the six APIs to include a condition that checks the selected approach (CCA or DCA) when the selected approach is CCA, the six APIs work as in the original \ourtool{}.
For instance, \texttt{DLS\_StartChunk} calls either \texttt{DLS\_StartChunk\_Centralized} or \texttt{DLS\_StartChunk\_Decentralized} based on the selected approach. 
\texttt{DLS\_StartChunk\_Centralized}  is a function that wraps the original CCA of \ourtool{}, while  \texttt{DLS\_StartChunk\_Decentralized} provides the newly added functionality that supports DCA.

%% file: 6.tex
\section{Performance Evaluation and Discussion }
\label{sec:ED}
Two \mbox{computationally-intensive} parallel applications are considered in this study to assess the performance potential of the proposed DCA. 
The first application, called PSIA~\cite{PSIA}, uses a parallel version of the \mbox{well-known} \mbox{spin-image} algorithm (SIA)~\cite{sia}.
SIA converts a 3D object into a set of 2D images. 
The generated 2D images can be used as descriptive features of the 3D object. 
As shown in Listing~\ref{algo:psia}, a single loop dominates the performance of PSIA.
	\begin{algorithm}[!b]
		\SetKwInOut{Input}{Inputs}
		\SetKwInOut{Output}{Output}
		spinImagesKernel (W, B, S, OP, M)\;
		\Input{W: \mbox{image width}, B: \mbox{bin size}, S: \mbox{support angle}, \mbox{OP: list of 3D points}, \mbox{M: number of spin-images}}
		\Output{R: list of generated spin-images} 
		\ShowLn	 
		\For{  {\color{blue}i = 0 $\rightarrow$ M}}
		{ 
			\ShowLn	
			P = OP[i]\;
			\ShowLn	
			tempSpinImage[W, W]\;
			\For{j = 0 $\rightarrow$ $length(OP)$}
			{
				\ShowLn	
				X = OP[j]\;
				\ShowLn	
				$np_i$ = getNormalVector(P)\;
				\ShowLn	
				$np_j$ = getNormalVector(X)\;
				\If{acos($np_i \cdot np_j$) $\le S$}
				{
					\ShowLn	
					$k$  =  $\Bigg \lceil$ $\cfrac{W/2 - np_i \cdot (X-P) }{B}$ $\Bigg \rceil$\;
					\ShowLn	
					\vspace{0.2cm}
					$l$ =  $\Bigg \lceil$  $\cfrac{ \sqrt{||X-P||^2 - (np_i\cdot(X-P))^2} }{B}$  $\Bigg \rceil$\;
					\ShowLn	
					\If{0 $\le$ k $\textless$ W and 0 $\le$ l $\textless$ W}
					{ tempSpinImage[k, l]++\;	}
				}
			}
			\ShowLn	
			R.append(tempSpinImage)\;
		}
		\caption{Parallel \mbox{spin-image} calculations. The main loop is highlighted in the {\color{blue}blue} color.}
		\label{algo:psia}
	\end{algorithm}

The second application calculates the Mandelbrot set~\cite{mandelbrot1980fractal}. 
The Mandelbrot set is used to represent geometric shapes that have the \mbox{self-similarity} property at various scales. 
Studying such shapes is important and of interest in different domains, such as biology, medicine, and chemistry~\cite{mandelbrot}.
	\begin{algorithm}[!t]
		\SetKwInOut{Input}{Inputs}
		\SetKwInOut{Output}{Output}
		mandelbrotSetCalculations (W, T)\;
		\Input{W: \mbox{image width}, CT: \mbox{Conversion Threshold}}
		\Output{V: Visual representation of mandelbrot set calculations }  
		\For{ {\color{blue} counter = 0 $\rightarrow$ $W^{2}$}}
		{ 
			\ShowLn	
			$x$ = counter $/$ W\;
			\ShowLn	
			$y$ = counter $\mod$ W\;
			\ShowLn	
			$c$= complex(x\_min + x/W*(x\_max-x\_min) , y\_min + y/W*(y\_max-y\_min))\;
			\ShowLn	
			$z$ = complex(0,0) \;
			\ShowLn	
			\For{$k=0 \rightarrow$ $CT$ OR $|z| \textless 2.0$}
			{
				\ShowLn	
				$z = z^4 + c$\;
			} 
			\ShowLn	
			\eIf{  $k=CT$}
			{
				\ShowLn	
				set $V(x,y)$ to black\;
			}
			{
				\ShowLn	
				set $V(x,y)$ to blue\;
			}
		}
		\caption{Mandelbrot set calculations. The main loop is highlighted in the {\color{blue} blue} color.}
		\label{algo:mandel}
	\end{algorithm}	
As shown in Listings~\ref{algo:psia} and~\ref{algo:mandel}, both applications contain a single large parallel loop iterations that dominates their execution times. 
Dynamic and static distributions of the most \mbox{time-consuming} parallel loop across all processing elements may enhance applications’ performance.
Table~\ref{tab:character} summarizes the characteristics of the main loops of both applications.

\begin{table*}[!t]
	\centering
	\caption{Characteristics of the selected applications' main loop}
	\label{tab:character}
\begin{tabular}{l|l|l}
\multirow{2}{*}{\begin{tabular}[c]{@{}l@{}}\textbf{Application characteristics} \end{tabular}} & \multicolumn{2}{c}{\textbf{Application}} \\
& \multicolumn{1}{l|}{\textbf{PSIA}} & \textbf{Mandelbrot} \\ \hline
	Number of loop iterations & 262,144 & 262,144 \\ \hline
	Maximum iteration execution time (s) & 0.190161 & 0.06237 \\ \hline
	Minimum iteration execution time  (s)& 0.0345 & 0.000001 \\ \hline
	Average iteration execution time  (s)& 0.07298 & 0.01025 \\ \hline
	Standard deviation (s)&0.00885 & 0.0187 \\ \hline
	Coefficient of variation~(c.o.v.)& 0.256 & 1.824 \\ \hline
\end{tabular}
\end{table*}

The target experimental system  is called miniHPC\footnote{\url{https://hpc.dmi.unibas.ch/HPC/miniHPC.html}}.
It consists of 26 compute nodes that are actively used for research and educational purposes. 
In the present work, we use sixteen dual-socket nodes.
Each node has two sockets with Intel Xeon E5-2640 processors and 10~cores per socket.

For each of the two applications, we evaluate the performance of twelve different techniques with both chunk calculation approaches: DCA and CCA.
Table~\ref{tab:doe} shows our design of factorial experiments in which each experiment is repeated 20 times.
All applications are compiled without compiler optimization~(-O0) using the Intel compiler version 19.1.0.166.
The Intel MPI Library for Linux OS version 2019 (update 6) is used to execute both applications.
\begin{table}[!h]
	\caption{Design of factorial experiments}
	\label{tab:doe}
	\resizebox{\textwidth}{!}{
		\begin{tabular}{ll}
				\textbf{Factor}                                      & \textbf{Value}                                                                                                                                        \\ \hline
				\multirow{2}{*}{Application}                & \begin{tabular}[c]{@{}l@{}}PSIA\\ \#spin-images= 262,144 (Total number of loop iterations to be scheduled)\\ image\_size= 5*5 \\ bin\_size=0.01\\ support\_angle= 0.5\end{tabular} \\ \cline{2-2} 
				& \begin{tabular}[c]{@{}l@{}}Mandelbrot\\ image\_size=512*512 (Total number of loop iterations to be scheduled)\\ conversion\_threshold= 1,000,000\end{tabular}                                      \\ \hline
				\multirow{1}{*}{Chunk calculation approach} & CCA and DCA\\ \hline                                                             
				Scheduling techniques                       & \begin{tabular}[c]{@{}l@{}}STATIC, FSC, GSS, TAP, TSS, FAC, TFSS,\\  FISS, VISS, RND, AF, PLS\end{tabular}                                   \\ \hline
				System                                      & \begin{tabular}[c]{@{}l@{}}16 Xeon nodes and 16 MPI ranks per node\\ Total of 256 MPI ranks\end{tabular}                                    \\ \hline
				Injected delay                                       & 0, 10, and 100 microseconds                                                                                                              \\ \hline
				Experiment repetitions                                 & 20                                                                                                                                           \\ \hline
			\end{tabular}
		}
	\end{table}

In the present work, we evaluate the performance potential of DCA and CCA in three different scenarios.
These scenarios represent cases when a slowdown affects the PEs and results in slowing down the  chunk calculation.
In the first scenario, no delay is injected during the chunk calculation.
In the other two scenarios, a constant delay is injected in the chunk calculation.
the injected delay was 10 and 100 microseconds for these two scenarios, respectively.

%% file: 7.tex
\begin{figure*}[!h]
		\centering
		
		\captionsetup[subfigure]{justification=centering}
		\centering
		\begin{subfigure}{0.9\textwidth}
			\centering
			\includegraphics[clip,trim=0cm 0cm 0cm 0cm, width=\textwidth]{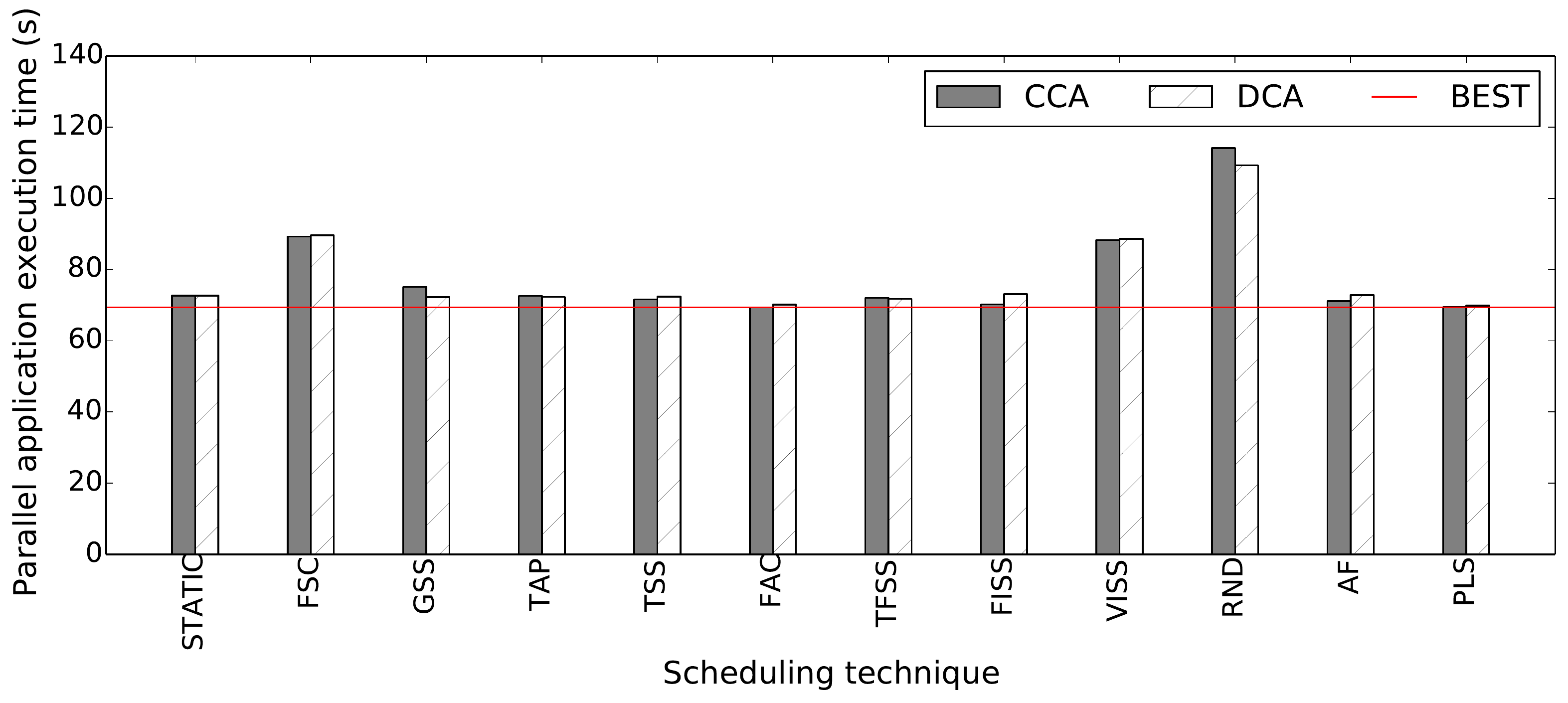}
			\subcaption{Without an injected delay}
			\label{fig:p_s0}%
		\end{subfigure}%
		
		\begin{subfigure}{0.9\textwidth}
			\centering
			\includegraphics[clip,trim=0cm 0cm 0cm 0cm, width=\textwidth]{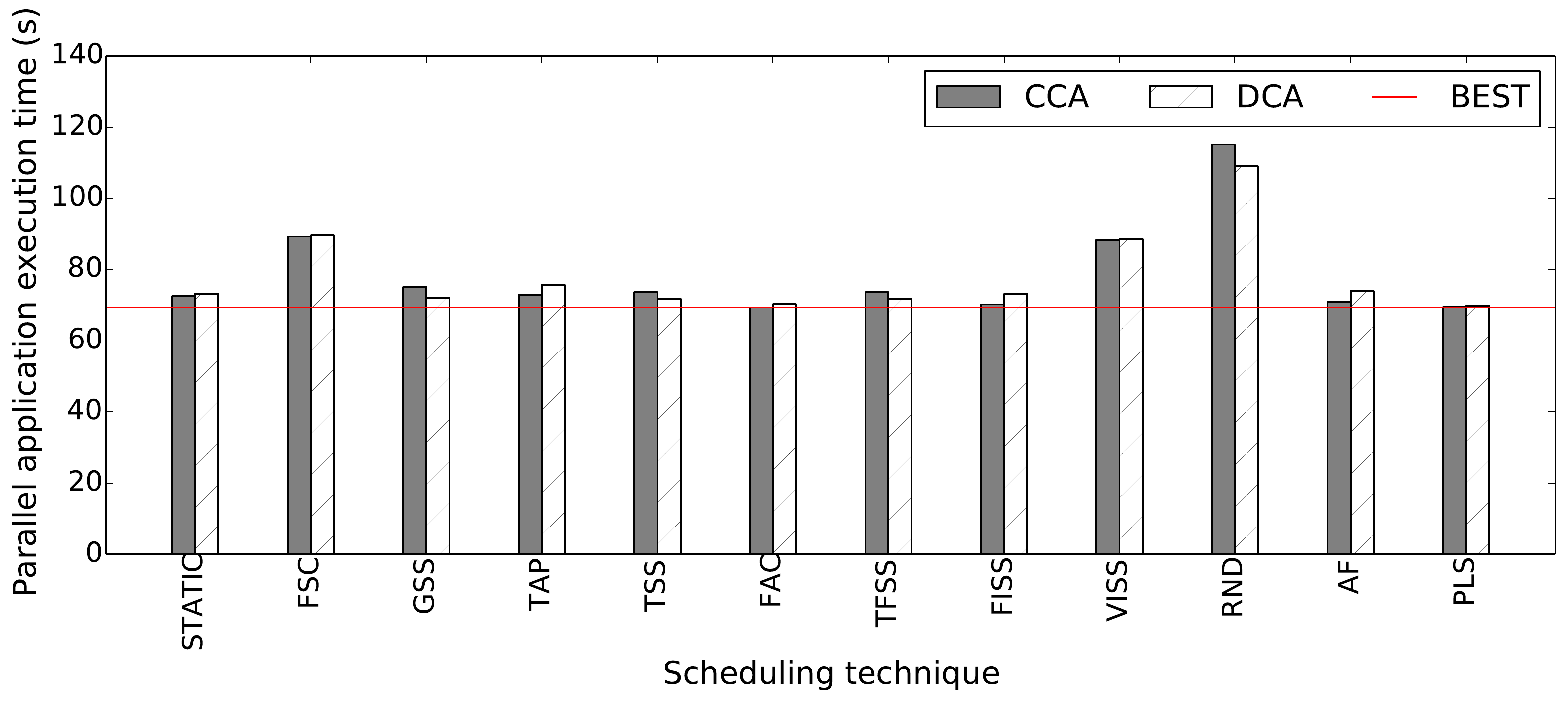}
			\subcaption{With low injected delay (10 microseconds)}
			\label{fig:p_s1}%
		\end{subfigure}%
		
		\begin{subfigure}{0.9\textwidth}
			\centering
			\includegraphics[clip,trim=0cm 0cm 0cm 0cm, width=\textwidth]{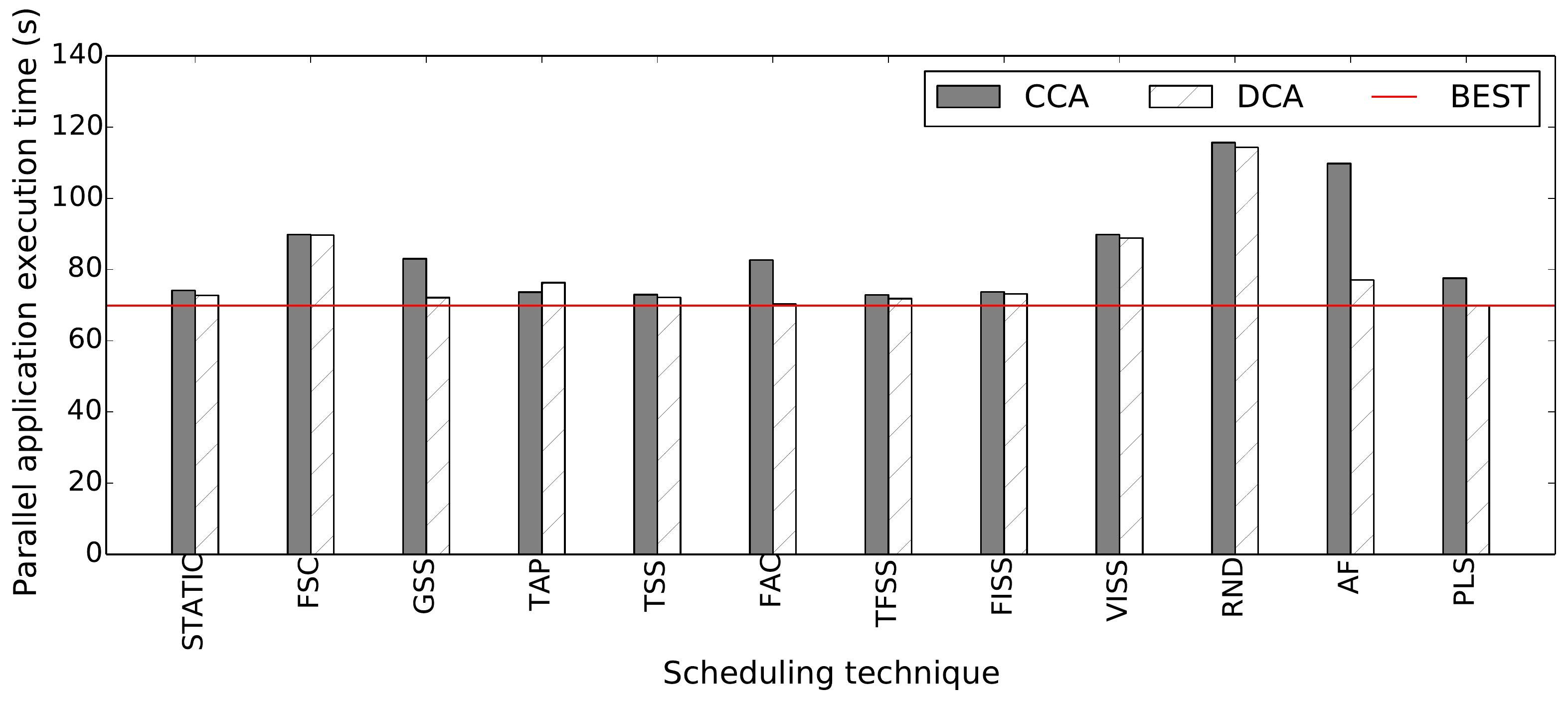}
			\subcaption{With severe injected delay (100 microseconds)}
			\label{fig:p_s2}%
		\end{subfigure}
		\caption{Parallel application execution time of PSIA in the three slowdown scenarios}
		\label{fig:PSIA}
	\end{figure*}
\begin{figure*}[!h]
	    	\centering
	    	
	    	\captionsetup[subfigure]{justification=centering}
	    	\centering
	    	\begin{subfigure}{0.9\textwidth}
	    		\centering
	    		\includegraphics[clip,trim=0cm 0cm 0cm 0cm, width=\textwidth]{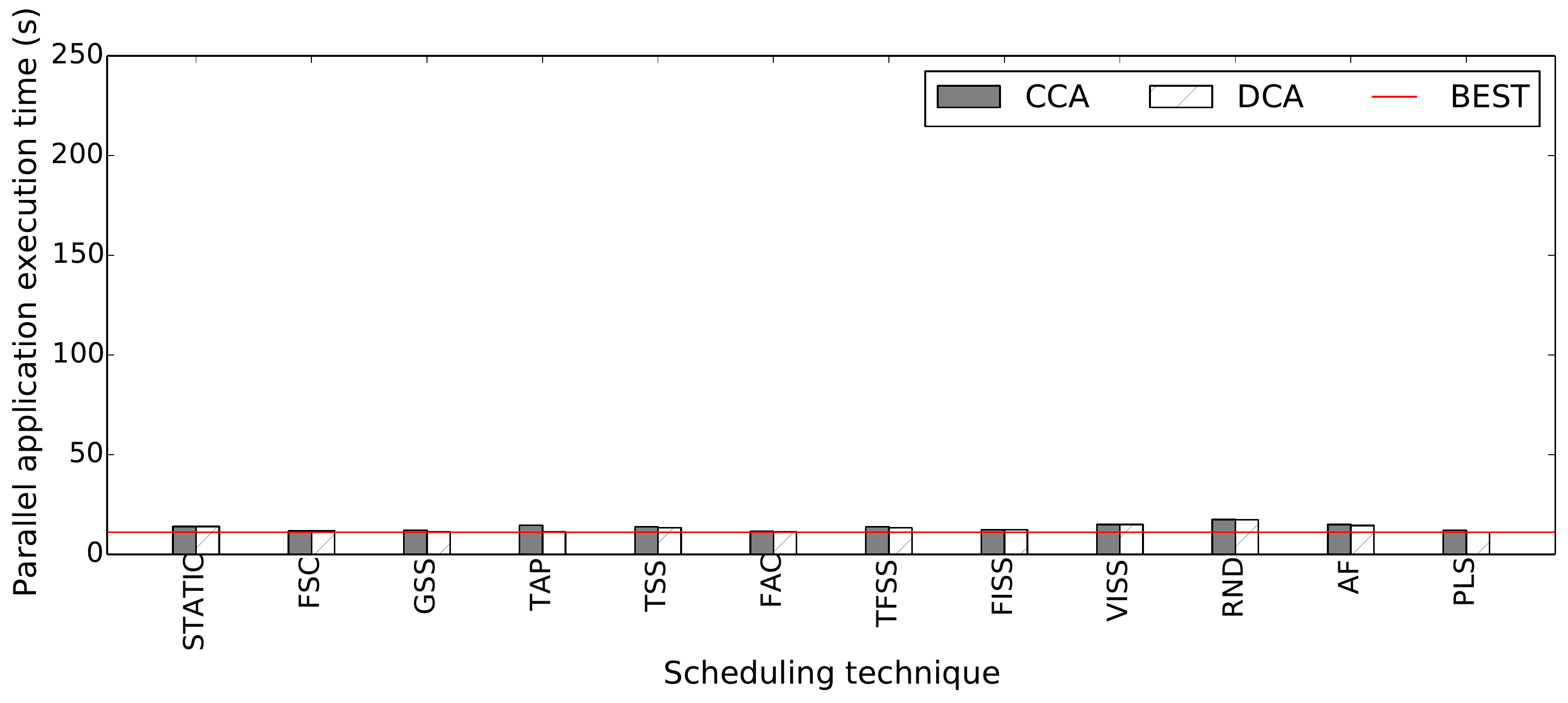}
	    		\subcaption{Without an injected delay}
	    		\label{fig:m_s0}%
	    	\end{subfigure}%
	    	
	    	\begin{subfigure}{0.9\textwidth}
	    		\centering
	    		\includegraphics[clip,trim=0cm 0cm 0cm 0cm, width=\textwidth]{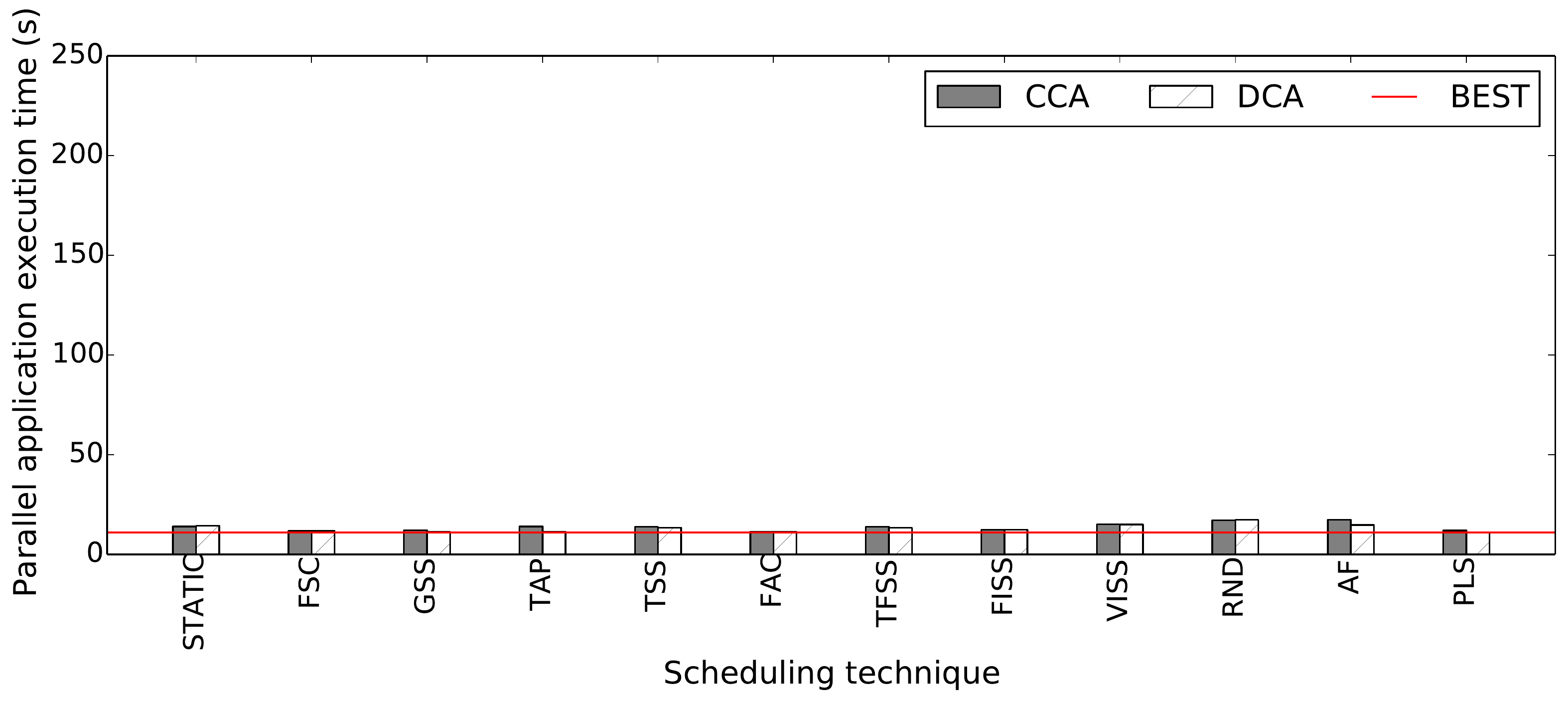}
	    		\subcaption{With low injected delay (10 microseconds) }
	    		\label{fig:m_s1}%
	    	\end{subfigure}%
	    	
	    	\begin{subfigure}{0.9\textwidth}
	    		\centering
	    		\includegraphics[clip,trim=0cm 0cm 0cm 0cm, width=\textwidth]{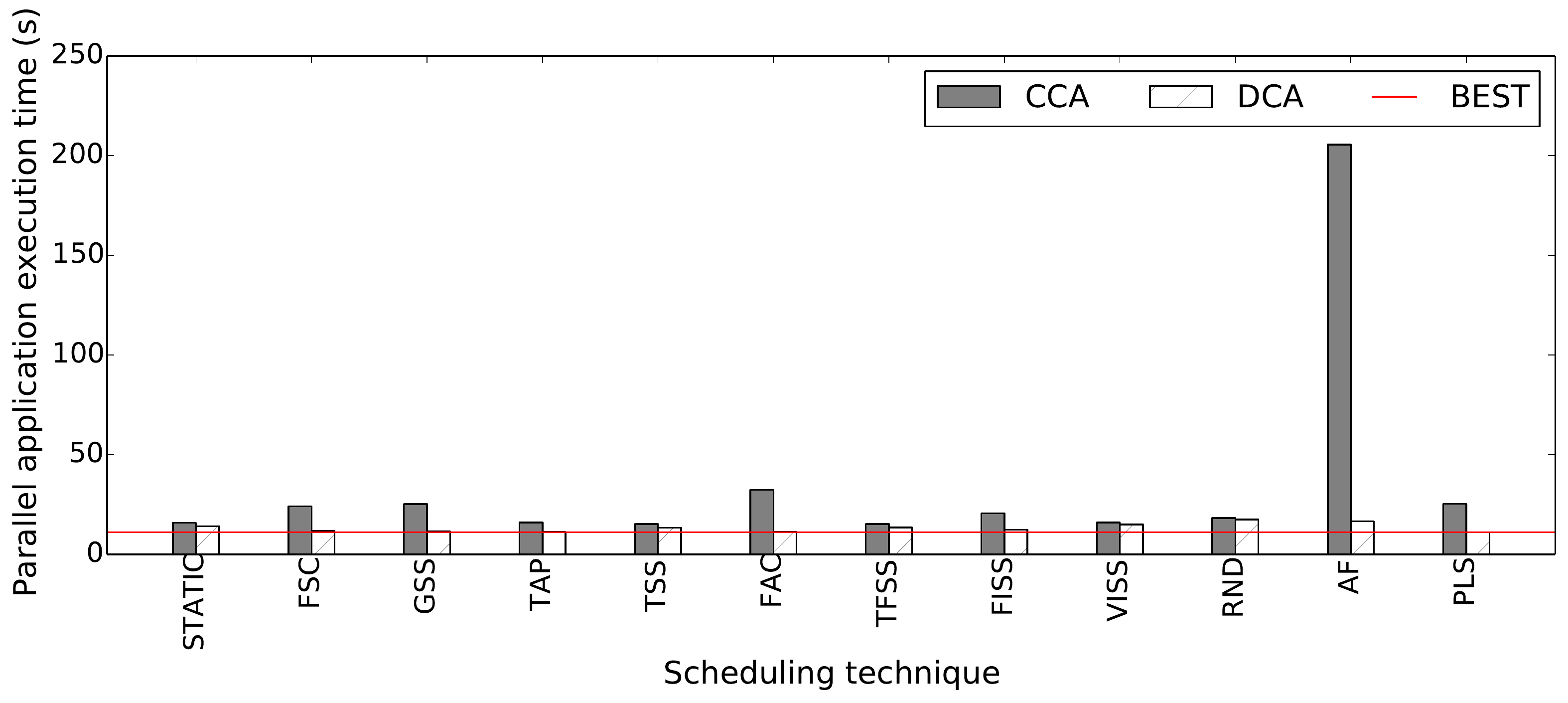}
	    		\subcaption{With severe injected delay (10 microseconds)}
	    		\label{fig:m_s2}%
	    	\end{subfigure}
	    	\caption{Parallel application execution time of Mandelbrot in the three slowdown scenarios}
	    	\label{fig:mandelbrot}
	    \end{figure*}

	Figures~\ref{fig:PSIA} and~\ref{fig:mandelbrot} show the performance of both CCA and DCA with different techniques for PSIA and Mandelbrot, respectively.
	As shown in Table~\ref{tab:character}, the coefficient of variation~(c.o.v.) for PSIA is significantly less than that of Mandelbrot. 
	This low c.o.v. indicates that PSIA has less load imbalance than Mandelbrot. 
	In Figure~\ref{fig:p_s0}, using CCA,  the parallel loop execution time $T_{loop}^{par}$ is 73.41 seconds with STATIC, while the best $T_{loop}^{par}$  is 69.37 with FAC.
	With FAC, the performance of PSIA is enhanced by 5.5\%. 
	Other techniques achieve comparable performance. 
	For instance, $T_{loop}^{par}$ is 69.53 seconds with PLS.
	In contrast, other techniques degrade the performance of PSIA. 
	GSS and RND degrade the PSIA performance by 2.7\% and 61.2\% compared STATIC.
	For the DCA, one can make the same observations regarding the best and the worst techniques.
	The CCA and DCA versions of all techniques are comparable to each other, i.e., the difference in performance ranges from 2\% to 3\%.
	
	Figures~\ref{fig:p_s1}  and~\ref{fig:p_s2} show the  performance of both CCA and DCA with different techniques for PSIA when the injected delay is 10 and 100 microseconds, respectively.
	In Figure~\ref{fig:p_s1}, one can notice that when the injected delay is 10 microseconds, 
	the performance differences between CCA and DCA with all techniques are in the range of~2\% to 3\%.
	Considering the variation in $T_{loop}^{par}$ of the 20 repetitions of each experiment, one observes that both approaches still have a comparable performance.
	For the largest injected delay, the DLS techniques implemented with CCA are more sensitive than the DLS techniques implemented with DCA (see Figure~\ref{fig:p_s2}).
	For Mandelbrot, one can notice the same behavior, i.e.,  when there is no injected delay or when the inject delay is 10  microseconds, 
	the performance differences between CCA and DCA with all techniques are minor (see Figures~\ref{fig:m_s0} and~\ref{fig:m_s1}).
	In contrast, Figure~\ref{fig:m_s2} shows that the DCA version of all the DLS techniques is more capable of maintaining its performance than the CCA version.

    Another interesting observation is the extreme poor performance of AF with CCA (see Figure~\ref{fig:m_s2}).
    AF is an adaptive technique, and it accounts for all sources of load imbalance that  affect applications during the execution.
    However, AF only considers $mu_{pi}$ and $\sigma_{pi}$. 
    Since we inject the delay in the chunk calculation function, AF cannot account for such a delay, and it works similarly to the case of no injected delay.
    Considering  the characteristics of the Mandelbrot application, the majority of the AF chunks are equal to 1 loop iterations.
    This fine chunk size leads to an increased number of chunks, i.e., the performance significantly decreased because the injected delay is proportional with the total number of chunks.
    For PSIA, the corresponding AF implementation (with CCA)  does not have the same extreme poor performance (see Figure~\ref{fig:p_s2}) because the AF chunk sizes in the case of PSIA are larger than the chunk sizes in the case of Mandelbrot.

%% file: 8.tex
\section{Conclusion and Future Work} 
\label{sec:conclusion}
In the present work, we studied how the distributed chunk calculation approach~(DCA)~\cite{DCA} can be applied to different categories of DLS techniques including DLS techniques that have fixed, decreasing, increasing, and irregular chunk size patterns.
The mathematical formula of the chunk calculation of any DLS technique can  either be straightforward or recursive.
The DCA requires that the mathematical formula of the chunk calculation be straightforward.
When one of the selected DLS techniques employs a recursive chunk calculation formula, we showed the mathematical transformations required to convert it into a straightforward formula.

By implementing the DCA in an \mbox{MPI-based} library called \ourtool{}~\cite{mohammed2019approach,mohammed2020simas}  using the \mbox{two-sided} MPI communication that is supported by all existing MPI runtime libraries, the present work answered the question:~\textit{Is DCA limited to specific MPI features?}

The present work showed the performance of CCA and DCA in three different slowdown scenarios.
In the first scenario, no delay was injected during the chunk calculation.
In the other two scenarios, a constant delay (small and large) was injected during the chunk calculation.
These scenarios represent cases when a slowdown affects the CPU and results in slowing down the  chunk calculation.
For these two scenarios, the injected delay was 0.00001 and 0.0001 seconds, respectively. 
For the large injected delay, the results showed that the DLS techniques implemented using the DCA were slightly affected by the injected delay.
This confirms the performance potential of the DCA~\cite{DCA}.
In a highly uncertain execution environment, when a slowdown affects the computational power of the coordinator (master), DCA is a better alternative to the CCA.

DCA incurs more communication messages than CCA, specifically  the message required to exchange scheduling data between the coordinator and the workers.
This increased number of messages could make DCA underperform CCA if the delay was injected during the chunk assignment  rather than the chunk calculation. 
Therefore, we plan to assess the performance of DCA with various communication slowdown scenarios.
Another future extension for \ourtool{} is to enable dynamic selection of the scheduling approach (DCA or CCA) that minimizes applications' execution time.